\documentclass[conference]{IEEEtran}
\IEEEoverridecommandlockouts
\usepackage[T1]{fontenc}
\usepackage[utf8]{inputenc}
\usepackage{booktabs}
\usepackage{tabularx}
\usepackage{array}
\usepackage{url}
\usepackage{balance}

\usepackage{cite}
\usepackage{amsmath,amssymb,amsfonts}
\usepackage{algorithmic}
\usepackage{graphicx}
\usepackage{textcomp}
\usepackage{xcolor}
\def\BibTeX{{\rm B\kern-.05em{\sc i\kern-.025em b}\kern-.08em
    T\kern-.1667em\lower.7ex\hbox{E}\kern-.125emX}}
\begin{document}

\title{Applying SHAPR in AI-Assisted Research Software Development: Lessons Learnt from Building a Share Trading System\\
%{\footnotesize \textsuperscript{*}Note: Sub-titles are not captured for https://ieeexplore.ieee.org  and should not be used}
}

\author{
\IEEEauthorblockN{Ka Ching Chan}
\IEEEauthorblockA{
School of Business, Law, Humanities and Pathways\\
University of Southern Queensland\\
Springfield Central, QLD 4300, Australia\\
Email: kc.chan@unisq.edu.au}
}

\iffalse
\author{\IEEEauthorblockN{1\textsuperscript{st} Given Name Surname}
\IEEEauthorblockA{\textit{dept. name of organization (of Aff.)} \\
\textit{name of organization (of Aff.)}\\
City, Country \\
email address or ORCID}
\and
\IEEEauthorblockN{2\textsuperscript{nd} Given Name Surname}
\IEEEauthorblockA{\textit{dept. name of organization (of Aff.)} \\
\textit{name of organization (of Aff.)}\\
City, Country \\
email address or ORCID}
\and
\IEEEauthorblockN{3\textsuperscript{rd} Given Name Surname}
\IEEEauthorblockA{\textit{dept. name of organization (of Aff.)} \\
\textit{name of organization (of Aff.)}\\
City, Country \\
email address or ORCID}
\and
\IEEEauthorblockN{4\textsuperscript{th} Given Name Surname}
\IEEEauthorblockA{\textit{dept. name of organization (of Aff.)} \\
\textit{name of organization (of Aff.)}\\
City, Country \\
email address or ORCID}
\and
\IEEEauthorblockN{5\textsuperscript{th} Given Name Surname}
\IEEEauthorblockA{\textit{dept. name of organization (of Aff.)} \\
\textit{name of organization (of Aff.)}\\
City, Country \\
email address or ORCID}
\and
\IEEEauthorblockN{6\textsuperscript{th} Given Name Surname}
\IEEEauthorblockA{\textit{dept. name of organization (of Aff.)} \\
\textit{name of organization (of Aff.)}\\
City, Country \\
email address or ORCID}
}
\fi

\maketitle

\begin{abstract}

Generative AI is changing how research software is developed, but rapid AI-assisted development can weaken continuity, traceability, and methodological clarity. SHAPR (Solo, Human-centred, AI-assisted PRactice) was proposed as a framework for structuring AI-assisted research software development. This paper presents a documented case of applying SHAPR to the development of a modular share trading system. From the outset, the project adopted a SHAPR-informed working configuration that shaped how interaction, implementation, and documentation were organised. Across iterative development cycles, the project generated a structured evidence base including reflection notes, development cycle review notes, source-of-truth documents, contracts, quick captures, workflow notes, and evolving code artefacts. The case showed that continuous documentation updates, supported by quick capture and AI-assisted refinement, helped maintain organised and usable project knowledge throughout development. Five recurring lessons were identified: contracts stabilised AI-assisted coding, a maintained source-of-truth layer improved coherence, cycle-boundary snapshots strengthened continuity, code and documentation co-evolved through quick capture and iterative refinement, and environment setup itself contributed to knowledge generation. The case also illustrates a practical SHAPR operating configuration in which a ChatGPT Project and cycle-specific chats supported interaction, reasoning, summarisation, and coding collaboration, PyCharm supported artefact implementation, and Obsidian supported external working memory, structured documentation, reflection, continuity, and repository-oriented note organisation, while remaining consistent with SHAPR’s tool-agnostic principle. The paper contributes practical guidance and good practices for researchers conducting AI-assisted research software development.

\end{abstract}

\begin{IEEEkeywords}
SHAPR, AI-assisted research software development, human-centred AI, traceability, agile workflow, personalisation, share trading system
\end{IEEEkeywords}

\section{Introduction}

Generative artificial intelligence is rapidly reshaping how research software is conceived, developed, and refined. Researchers can now use large language models and related tools to generate code, suggest architectural options, support debugging, summarise evolving work, and accelerate exploratory development. These capabilities are particularly attractive in software-intensive research projects, where solo researchers often need to move quickly across problem framing, implementation, evaluation, and reflection. However, faster development does not automatically produce more traceable, coherent, or methodologically disciplined research practice.

In AI-assisted development contexts, several practical risks arise repeatedly. These include naming drift, scope drift, fragmented documentation, weak continuity across sessions, and difficulty linking conversational exploration to stable artefacts and research evidence. These risks are especially important in research settings, where the value of the software artefact is tied not only to functionality, but also to the visibility of decisions, the traceability of changes, and the researcher’s ability to explain how knowledge emerged through iterative development \cite{Stodden2010}. A practical framework is therefore needed to help researchers preserve human judgement while still benefiting from AI assistance \cite{Amershi2019,Shneiderman2022}.

SHAPR (Solo, Human-centred, AI-assisted PRactice) was proposed to address this need by structuring AI-assisted research software development around human-centred governance, iterative development cycles, artefact-centred evidence, and reflective learning \cite{chan2026shapr,chan2026operational}. Earlier SHAPR work established the conceptual and operational foundations of the framework. The next step, however, is practical dissemination: researchers need to see how SHAPR can actually be enacted in a real project, what kinds of records are useful, what difficulties emerge in practice, and what lessons can be transferred to future work.

This paper contributes to that practical step by examining a documented case in which SHAPR was applied to the development of a modular share trading system. The value of the case lies not primarily in the trading domain itself, but in the richness of the development evidence generated across development cycles. The project produced reflection notes, cycle review notes, source-of-truth documents, quick captures, contracts, workflow records, and evolving code artefacts, creating a substantial basis for identifying recurring patterns in practice. Rather than presenting the case mainly as a domain artefact, this paper focuses on the lessons learnt from applying SHAPR in a real software-intensive context.

A distinctive feature of the case is that documentation was updated continuously rather than treated as a delayed reporting task. Quick captures preserved emerging observations and decisions with low friction, and these records were later refined with AI assistance into more organised and reusable notes. This helped keep project knowledge visible and usable across development cycles without placing excessive documentation burden on the researcher. The case also highlights the importance of personalisation: SHAPR does not prescribe a fixed documentation regime, but supports the collaborative design of a working configuration suited to the researcher, the project, and the chosen tools. At project initiation, the two prior SHAPR arXiv papers were uploaded to ChatGPT, allowing the framework itself to help shape the project environment \cite{chan2026shapr,chan2026operational}.

Five lessons are emphasised in this paper: contracts and control artefacts stabilised AI-assisted coding; a maintained source-of-truth layer improved coherence and implementation quality; cycle-boundary snapshots and handoff artefacts strengthened continuity and traceability; code and documentation co-evolved through quick capture and refinement; and setting up a SHAPR-compliant environment was itself knowledge-generating. The case also provides a practical example of SHAPR enacted through linked workspaces: a ChatGPT Project with cycle-specific chats for interaction and collaboration, PyCharm for artefact implementation, and Obsidian for external working memory, documentation, reflection, and project continuity.

The paper therefore makes two compact contributions. It presents a real case of SHAPR in practice through the development of a modular share trading system, and it distils that case into practical guidance and good practices that other researchers can adapt in their own AI-assisted research software projects. By focusing on lessons learnt rather than on domain-specific share trading outcomes, the paper aims to support wider SHAPR adoption in software-intensive research.

\section{SHAPR and the Share Trading System Case}

\subsection{SHAPR in Brief}

SHAPR, or Solo, Human-centred, AI-assisted PRactice, is a framework for conducting AI-assisted research software development in a structured and traceable way. It places human judgement, accountability, and reflective learning at the centre of development practice, while treating AI as a collaborator that can support coding, exploration, summarisation, and documentation rather than as an autonomous decision-maker.

A key feature of SHAPR is its emphasis on iterative development as a knowledge-generating process \cite{Sein2011}. Development activities are not treated only as technical implementation; they also generate research evidence through artefacts, decisions, reflections, reviews, and changing project records. This makes documentation and reflective control central rather than optional.

SHAPR is also deliberately tool-agnostic. Rather than prescribing a fixed toolset or working style, it allows researchers to choose tools, rhythms, and levels of documentation suited to their own context, provided that human-centred governance, iterative discipline, and evidence visibility are preserved. This makes SHAPR compatible with personalised solo workflows and agile-style iteration.

This case is especially relevant to the operational SHAPR view in which practice is supported by both an interaction workspace and a repository workspace \cite{chan2026operational}. It also aligns with the moderate-AI involvement configuration described in earlier SHAPR work, where AI supports reasoning, summarisation, documentation, and implementation assistance, while the researcher remains responsible for direction, interpretation, and final decisions.

The present paper does not restate the SHAPR framework in full. Instead, it builds on earlier conceptual and operational SHAPR work by focusing on what can be learnt from applying the framework in a real project.

\subsection{Case Context}

The case examined in this paper involved the iterative development of a modular share trading system intended to support software-based investigation of indicators, derived signals, strategies, and backtesting pathways. The project progressed through documented development cycles covering environment setup and design formation (Development Cycle 0), baseline coding (Development Cycle 1), backend stabilisation (Development Cycle 2), indicator development (Development Cycle 3), and derived-signal work (Development Cycle 4). Across these cycles, the case generated a substantial set of records that went beyond source code alone. The present paper focuses on Development Cycles 0 to 4 as a bounded case for analysis, while the broader project is expected to continue through later cycles beyond the scope of this paper.

The value of the case for this paper lies in three characteristics. First, it was sufficiently complex to create real development tensions involving structure, naming, control, documentation, and continuity. Second, it was deliberately developed in a modular way, allowing specific lessons to emerge across staged implementation. Third, it produced rich supporting materials, including cycle notes, cycle review notes, decisions, source-of-truth records, quick captures, contracts, and workflow reflections. These materials make the case useful not only as a software artefact, but also as an evidence base for deriving practical lessons about SHAPR enactment.

The project also provides a concrete example of SHAPR’s practical tool-agnostic principle. In this case, SHAPR was enacted through linked workspaces in which a ChatGPT Project supported project-level interaction, multiple chats served as cycle-specific workspaces, PyCharm supported implementation, and Obsidian supported source-of-truth notes, reflections, contracts, cycle records, quick captures, and cycle continuity. The significance of this pattern is not that these tools are mandatory, but that it demonstrates how complementary environments can support code development and reflective control together in a practical SHAPR configuration.

\section{Practical SHAPR Enactment in the Case}

\subsection{Linked Workspaces and SHAPR Compliance}

A major practical insight from the case was that SHAPR compliance was supported not by a single tool, but by a linked set of workspaces. At project initiation, the two prior SHAPR arXiv papers were uploaded to ChatGPT \cite{chan2026shapr,chan2026operational}. This allowed the framework itself to help shape the practical working environment for the project and supported the identification of a configuration suited to the researcher’s working style and project needs.

In this implementation, a ChatGPT Project provided the overarching interaction environment, while multiple chats were used as cycle-specific workspaces. This made it possible to align each chat with a development cycle while preserving continuity at the broader project level. Within these interaction workspaces, ChatGPT Thinking supported reasoning, option generation, summarisation, documentation refinement, and close coding collaboration. AI was therefore not limited to code suggestion alone; it also supported problem framing, note refinement, and practical workflow decisions. Final decisions, however, remained with the researcher.

PyCharm functioned as the implementation workspace, supporting code organisation, modular development, execution, and debugging. Obsidian functioned as the repository and documentation workspace, supporting source-of-truth notes, contracts, decisions, cycle records, reflections, quick captures, and snapshots used to preserve continuity across development cycles and chat-based work. Together, these workspaces formed a practical SHAPR operating configuration that supported implementation control, reflective visibility, and continuity without forcing all activities into a single environment.

This arrangement is consistent with the moderate-AI involvement pattern described in earlier SHAPR work \cite{chan2026operational}: AI remained closely involved in reasoning, summarisation, documentation refinement, and coding support, but the researcher retained authority over requirements, acceptance, interpretation, and final decisions. The significance of this configuration is not that these exact tools are mandatory, but that it demonstrates an adoptable pattern for enacting SHAPR in practice.

\begin{figure}[t]
\centering
%\fbox{\rule{0pt}{1.4in}\rule{0.95\columnwidth}{0pt}}
\includegraphics[width=0.9\columnwidth,trim={0 5mm 0 3mm},clip]{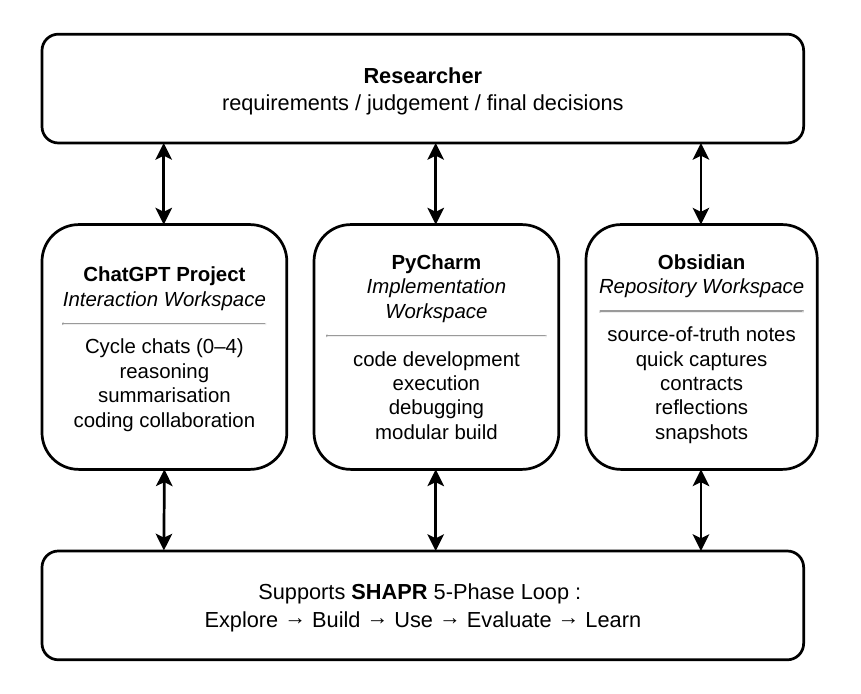}
\caption{Practical SHAPR operating configuration used in the case. The researcher remained central while working across a ChatGPT Project as the interaction workspace, PyCharm as the implementation workspace, and Obsidian as the repository workspace. Together, these workspaces supported the recurring SHAPR five-phase loop of Explore, Build, Use, Evaluate, and Learn.}
\label{fig:environment}
\end{figure}

\subsection{Documentation Structure}

Another important part of the case was the practical organisation of documentation. Rather than treating notes as a flat collection, the project developed a structured working environment with distinguishable roles, including system notes for architecture and module design, contract notes for function rules, data structures, and naming conventions, control notes for coding protocol, decisions, and workflow, working notes for current state and active tasks, cycle records and cycle review notes, and snapshots and handoff artefacts at cycle boundaries.

This structure reduced ambiguity about note roles and supported stronger project coherence. A source-of-truth layer could guide active implementation, while other notes captured more specific decisions, module details, and historical records. It also supported later synthesis by ensuring that recurring issues and lessons were not hidden in unstructured conversational traces.

The code and documentation supporting the findings of this study are available in Zenodo \cite{chan2026shapr_repo}. This repository is intended not only to preserve evidence from the case, but also to support adoption by allowing future researchers to inspect, adapt, and initiate their own SHAPR-based software projects using a documented example.

\subsection{Continuous Updates, Quick Capture, and AI-Assisted Refinement}

A distinctive practice in the case was the ongoing updating of documentation throughout development. Rather than postponing note maintenance until the end of a development cycle, the project used quick captures to preserve emerging observations, ideas, design issues, and implementation concerns as they arose. These quick captures lowered the cost of documentation and reduced information loss during active work.

Across the case, development activity was interpreted through the recurring SHAPR five-phase loop of \textit{Explore} $\rightarrow$ \textit{Build} $\rightarrow$ \textit{Use} $\rightarrow$ \textit{Evaluate} $\rightarrow$ \textit{Learn}. Quick captures were especially valuable because they allowed the researcher to record observations, design ideas, issues, and emerging lessons with low friction as they arose within and across these phases. This reduced the risk that important insights would be lost during active work, while also allowing the researcher to remain focused on coding when appropriate.

AI assistance then helped refine these captures into more organised and usable forms. In this case, ChatGPT supported the conversion of partial notes into more structured documentation, strengthening the connection between the interaction workspace and the repository workspace. This did not replace human judgement, but supported the transformation of fragmented observations into reusable records.

Over successive development cycles, these captures, refinements, review notes, and source-of-truth updates accumulated into a substantial and readily available body of project knowledge. By the end of the case, the notes were valuable not only as a memory aid, but also as a structured evidence base supporting reflection, lesson extraction, and practical guidance. This suggests that SHAPR can benefit from a documentation rhythm that is lightweight, iterative, and cumulative: small captures made during active work can later become an important resource for continuity, analysis, and knowledge generation. The documentation preserved in the shared Zenodo repository \cite{chan2026shapr_repo} was produced through this same process of quick capture, iterative refinement, cycle review, and source-of-truth updating, rather than being assembled retrospectively after development. This increases the practical value of the repository for future SHAPR users, because the records reflect the evolving development process itself.

\begin{figure}[t]
\centering
%\fbox{\rule{0pt}{1.25in}\rule{0.95\columnwidth}{0pt}}
\includegraphics[width=0.8\columnwidth,trim={0 6mm 0 0mm},clip]{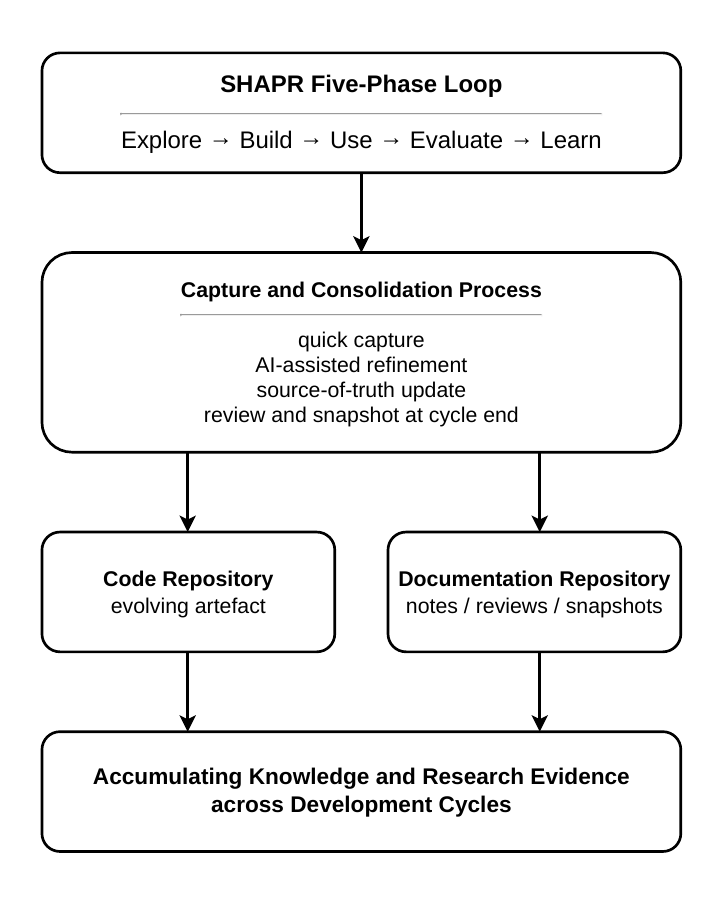}
\caption{General SHAPR process used within each development cycle. The SHAPR five-phase loop of Explore, Build, Use, Evaluate, and Learn was supported by a capture and consolidation process that updated both the code and documentation repositories, allowing SHAPR records to accumulate across development cycles.}
\label{fig:transition}
\end{figure}

\subsection{Cycle-to-Cycle Transition}

The case also demonstrated a practical pattern for moving from one development cycle to the next. In simplified form, the pattern was:
\begin{enumerate}
    \item active implementation and note updating occur during the current development cycle,
    \item review and reflection consolidate what was learnt,
    \item source-of-truth notes are updated,
    \item a cycle review or handoff record is preserved,
    \item snapshots or equivalent artefacts capture the stabilised state, and
    \item the next cycle begins from that preserved state, often in a new cycle-specific chat workspace within the broader project environment.
\end{enumerate}

This transition logic matters because it reduces dependence on informal continuity and strengthens both operational and methodological control. In SHAPR terms, development cycle transitions are not merely administrative boundaries; they are opportunities to stabilise artefacts, preserve evidence, and convert development activity into reusable learning.

\begin{table*}[t]
\caption{Lessons Learnt, Main Evidence, and Practical Implications}
\label{tab:lessons}
\centering
\footnotesize
\begin{tabularx}{\textwidth}{>{\raggedright\arraybackslash}p{0.19\textwidth} >{\raggedright\arraybackslash}p{0.42\textwidth} >{\raggedright\arraybackslash}X}
\toprule
\textbf{Lesson} & \textbf{Main evidence} & \textbf{Practical implication} \\
\midrule
Contracts stabilise AI-assisted coding & Cycle and review notes highlighted naming drift, interface drift, and structural inconsistency. Contracts, naming rules, and coding protocol notes were introduced early and refined later. & Define control artefacts early, including function contracts, data structures, naming rules, and code maps or equivalent interface records. \\
\addlinespace
A maintained source-of-truth layer improves coherence & Development became more fragile when active notes were unclear or outdated. Current-state and source-of-truth notes helped align project direction with code evolution. & Maintain one active source-of-truth layer recording project state, current direction, key decisions, and immediate next steps. \\
\addlinespace
Snapshots strengthen continuity and traceability & Snapshot workflow and handoff practices improved continuity by preserving a documented state at the end of each cycle. & Preserve cycle boundaries through snapshots, handoff artefacts, or equivalent records carrying artefact state and development rationale. \\
\addlinespace
Code and documentation should co-evolve through quick capture and refinement & Review notes, quick captures, and implementation revisions showed that low-friction capture preserved emerging observations, while later refinement kept notes usable and aligned with code. & Use quick capture during active work, then refine notes iteratively so documentation stays current without distracting from coding. \\
\addlinespace
Environment setup is knowledge-generating & Cycle 0 materials showed that note structure, contracts, workflow setup, and environment design shaped later coding quality and reflection. & Treat environment design, documentation workflow, and control-layer setup as part of the research process rather than as overhead. \\
\bottomrule
\end{tabularx}
\end{table*}

\section{Key Lessons Learnt from the Case}

Table~\ref{tab:lessons} summarises the main lessons derived from the documented case materials, including development cycle records, review notes, source-of-truth documents, quick captures, and supporting artefacts archived in the project repository \cite{chan2026shapr_repo}.

The case produced five recurring lessons:
\begin{enumerate}
    \item contracts and control artefacts stabilised AI-assisted coding;
    \item a maintained source-of-truth layer improved coherence and implementation quality;
    \item cycle-boundary snapshots and handoff artefacts strengthened continuity and traceability;
    \item code and documentation co-evolved through quick capture and iterative refinement; and
    \item setting up a SHAPR-compliant environment was itself knowledge-generating.
\end{enumerate}

Taken together, these lessons suggest that SHAPR is most effective when implementation control, documentation discipline, and development cycle continuity are treated as integrated parts of the research process rather than as secondary support activities. A particularly important practical finding was that quick capture and ongoing refinement reduced documentation burden while keeping project knowledge visible and usable. This allowed the researcher to remain focused on coding when appropriate, without losing reflective visibility or continuity across development cycles.

The lesson set also reinforces the practical value of SHAPR’s personalised and tool-agnostic stance. Different researchers may prefer different tools, rhythms, and note structures, but the case indicates that effectiveness depends less on one fixed configuration than on preserving a workable balance between interaction, implementation, documentation, and reflection. In this sense, the lessons reported here are better understood as transferable good practices than as rigid rules, and can therefore be adapted to different solo AI-assisted research contexts.

\section{Good Practices for Applying SHAPR}

The lessons from the case can be translated into a compact set of good practices for future SHAPR users. These include defining stabilising control artefacts early, maintaining one active source-of-truth layer, preserving development cycle boundaries through snapshots or equivalent handoff artefacts, updating documentation continuously alongside implementation, and treating environment and workflow design as part of the research process itself. Quick captures are especially useful because they preserve emerging information with low friction, while later AI-assisted refinement can organise those records into more reusable forms.

A further implication of the case is that SHAPR supports personalisation. Solo researchers differ in how they organise notes, how frequently they update documentation, and how much structure they prefer. SHAPR does not require identical practices across users. Instead, it enables researchers to choose the level of documentation detail, update rhythm, and tool combination that best supports their own working style, provided that evidence visibility, continuity, and human-centred control are maintained. In this sense, SHAPR aligns naturally with agile-style work: iterative, adaptive, and personalised, but still governed by deliberate checkpoints and reflective control.

A compact checklist derived from the case is as follows:
\begin{itemize}
    \item[$\square$] Is there one clearly identified active source-of-truth note?
    \item[$\square$] Are core interfaces, structures, and naming rules explicitly documented?
    \item[$\square$] Are quick captures or equivalent low-friction records being used during active work?
    \item[$\square$] Are documentation updates happening continuously alongside implementation?
    \item[$\square$] Is each development cycle concluded with a review and a preserved snapshot or handoff record?
    \item[$\square$] Can another researcher reconstruct how the project evolved across development cycles?
    \item[$\square$] Does the current tool and note configuration support both implementation and reflective continuity?
\end{itemize}

These checks are intentionally lightweight. Their purpose is not to enforce a rigid process, but to help future SHAPR users preserve continuity, coherence, and traceability while adapting the framework to their own preferred working style.

\section{Conclusion}

This paper presented a documented case of applying SHAPR to the development of a modular share trading system and used that case to derive practical lessons for future AI-assisted research software development projects. The main contribution of the paper lies not in the trading artefact alone, but in the practical guidance that emerged from case evidence including reflections, review notes, source-of-truth documents, quick captures, contracts, and workflow records.

Five lessons were especially clear. Explicit contracts and control artefacts helped stabilise AI-assisted coding. A maintained source-of-truth layer improved coherence and implementation quality. Cycle-boundary snapshots and handoff artefacts strengthened continuity and traceability across development cycles. Code and documentation needed to co-evolve through quick capture and iterative refinement. Setting up a SHAPR-compliant environment was itself a meaningful design and learning activity.

Taken together, these findings suggest that SHAPR is not only a conceptual and operational framework, but also a practical and teachable way of conducting AI-assisted research software development. The case illustrated a linked SHAPR operating configuration in which a ChatGPT Project and cycle-specific chats supported interaction, reasoning, summarisation, and coding collaboration, PyCharm supported artefact implementation, and Obsidian supported external working memory, documentation, reflection, continuity, and repository organisation, while remaining consistent with SHAPR’s tool-agnostic principle.

Future work can extend this study through fuller development of the share trading system, systematic experimentation with different AI-involvement configurations within SHAPR, and the development of a collaborative repository and record-management system to support SHAPR-based software development in practice.

%\section*{References}
\balance

\end{document}